\documentclass[12pt]{article}
\usepackage{graphicx}
\usepackage[english]{babel}
\usepackage[latin1]{inputenc}
\usepackage{mathrsfs}
\usepackage{amsfonts}
\usepackage{amsmath}
\usepackage{amssymb}

\ifx\pdfoutput\undefined
\usepackage[hypertex]{hyperref}
\else
\usepackage[pdftex,hypertexnames=false]{hyperref}
\fi

\begin{document}

\title{Semi-explicit Parareal method based on convergence acceleration technique}
\author{Lo\"ic MICHEL}
\maketitle
\begin{abstract}
The Parareal algorithm is used to solve time-dependent problems considering multiple solvers that may work in parallel. The key feature is a initial rough approximation of the solution that is iteratively refined by the parallel solvers.
We report a derivation of the Parareal method that uses a convergence acceleration technique to improve the accuracy of the solution. Our approach uses firstly an explicit ODE solver to perform the parallel computations 
with different time-steps and then, a decomposition of the solution into specific convergent series, based on an extrapolation method, allows to refine the precision of the solution. Our proposed method exploits basic 
explicit integration methods, such as for example the explicit Euler scheme, in order to preserve the simplicity of the global parallel algorithm. The first part of the paper outlines the proposed method applied to the simple 
explicit Euler scheme and then the derivation of the classical Parareal algorithm is discussed and illustrated with numerical examples.
\end{abstract}
\section{Introduction}
\label{intro}
Despite the lack of accuracy that prevent it from being used as an ODE solver, the Euler method is known to be very easy to implement \cite{Yang}. Many efficient methods have been successfully implemented to solve ODEs in a parallel way. 
For this purpose, the Parareal algorithm \cite{Lions} introduces the possibility of parallelizing the computations requested to solve efficiently ODEs using e.g. an Euler scheme. This method uses a predictor-corrector algorithm, 
applied to an implicit Euler scheme, and allows faster simulations, especially if parallel computing is available \cite{Maday}. Further analysis and stability proofs can be found e.g. in \cite{Maday} \cite{Maday2} \cite{Staff} \cite{Vand} \cite{Vand2}. 
The Parareal method uses originally coarse and fine implicit propagators to evaluate iteratively the solution of the considered ODEs, whose precision increases according to the iterations.

To extend the original properties, the proposed method uses, after the first iteration, an explicit Euler solver for both coarse and fine propagators in order to avoid difficulties related to matrix inversion and thus to keep
the advantages of the simplest Euler scheme. The fine propagator recomputes iteratively and corrects the solution using different time-steps, that can be extrapolated using a convergence acceleration method in order to reach a more 
accurate solution. The use of extrapolation algorithms has been already considered to improve the accuracy of the solution of ODEs considering the extrapolation of the expansion of the error relating to the powers of the time-step in the 
case of explicit and implicit numerical integration methods (e.g. \cite{Deu} \cite{Sandu}). Other derivations of the Parareal method, including e.g. a symplectic derivation \cite{Bal}, an adjoint-based approach \cite{Rao}, a relaxation-based 
derivation \cite{Liu} and a multiscale approach \cite{Legoll} have been also successfully applied. 

The paper is structured as follows. Section 2 presents the outline of the method applied to the explicit Euler scheme. Section 3 gives a brief overview of the acceleration algorithm used to increase the precision of the solution of ODEs regarding
the application to the explicit Euler scheme. Section 4 reviews the standard Parareal algorithm and describes the proposed derivation including some illustrations with numerical simulations. Some concluding remarks may be found in Section 5.
\section{Outline of the method}\label{sec:secI}
In this section, the proposed method is applied to the simple explicit Euler scheme before being applied to the Parareal algorithm using the same principle.

Consider an ordinary differential equation, eventually non-linear, such as:
\begin{equation}\label{eq:EDP_fond} 
\frac{\mathrm{d} \, y(t)}{\mathrm{d} \, t} = \mathbf{A}(t) y(t) + \mathbf{B}(t) u(t), \qquad t \in [ 0, \, T_f], \quad y(0) = y_0
\end{equation}
\noindent
The quantities $u$ and $y$ represent respectively the input and the solution of (\ref{eq:EDP_fond}), which can be a scalar or a vector. To solve (\ref{eq:EDP_fond}) and to compute an {\it initial} 
solution over $[ 0, \, T_f]$, we use the forward Euler method, for which we assume {\it initially} that the time-step $h_0$ is large and ensure the stability. The equation (\ref{eq:EDP_fond}) is rewritten in the discrete time-domain:
\begin{equation}
\frac{y^0 (t^0_{k_0+1}) - y^0 (t^0_{k_0})}{h_0} = \mathbf{A_{k_0}} y^0 (t^0_{k_0}) + \mathbf{B_{k_0}} u_{k_0}
\end{equation}
\noindent
whose solution $y^0$ verifies ($\mathbf{I}$ is the identity matrix whose size is compatible with the size of $\mathbf{A}$):
\begin{equation}\label{eq:EDP_disc}
y^0 (t^0_{k_0+1}) = (\mathbf{I} + \mathbf{A_{k_0}} h_0 ) y^0 (t^0_{k_0}) + \mathbf{B_{k_0}} h_{k_0} u_{k_0} 
\end{equation}
\noindent
The coefficient $h_0$ is the {\it initial} time-step such as the solution $y^0$ is calculated at each instant $t^0_{k_0} = k_0 \, h_0$, $k_0 \in \mathcal{K}_0 \subset \mathbb{N}$. $\mathbf{A}_{k_0}$ and $\mathbf{B}_{k_0}$ are 
"connected" to $\mathbf{A}(t)$ and $\mathbf{B}(t)$ and are described by specific relationships that depend on the smoothness of $\mathbf{A}(t)$ and $\mathbf{B}(t)$. In this paper, we consider $\mathbf{A}$ and $\mathbf{B}$ that do not 
depend on the time.  We call $\mathbf{I} + \mathbf{A_{k_0}} h_0$, the dynamic matrix of the explicit Euler scheme. The value $y_0$ is considered as the initial condition. 
\noindent
To solve accurately the explicit Euler scheme, one performs a first resolution using a high time-step $h_0$. Then, for each instant $k_0 h_0$ of the initial solution, the 
initial time-step $h_0$ is divided by a finite number $i \in \mathcal{I} \subset \mathbb{N}$ of subdivisions $\delta_i$ such as $h_i = h_0 / \delta_i$. Some power computations of ($\mathbf{I} + \mathbf{A_{k_0}} h_i$) give a sequence 
for which, the limit, and therefore, a good estimation of the true solution, is deduced using a convergence accelerator.
\noindent
We denote:
\begin{equation*}
\mathbf{\Psi^0} = ( y_0, \, y^0(t^0_1), \, y^0(t^0_2), \, \cdots, \, \underline{ y^0(t^0_{k_0}) }, \,  \cdots, \, y^0(T_f) )
\end{equation*}
\noindent
the series that contains the points of the calculated solution (with the time-step $h_0$) at each instant $t^0_{k_0} = k_0 h_0$.
\noindent
Consider now a smaller time-step $h_1  = h_0 / \delta_1 < h_0$, with $\delta_1 > 1$, for which we obtain the series of the calculated solutions:
\begin{equation*}
\mathbf{\Psi^1} = ( y_0, \, y^1(t^1_{\delta_1}), \, y^1(t^1_{2 \delta_1}), \, \cdots, \, \underline{ y^1(t^1_{k_0 \delta_1}) }, \,  \cdots, \, y^1(T_f) )
\end{equation*}
\noindent
at each instant $t^1_{k_1} = k_1 h_1 = k_1 h_0 / \delta_1, \, k_1 \in \mathcal{K}_1 \subseteq  \mathcal{K}_0 \subset \mathbb{N}$. We have thus $t^1_{k_0 \delta_1} = t^0_{k_0}$.
\noindent
Therefore, by induction, we can deduce that for any time-step $h_{i} = h_0 / \delta_i < h_{i-1}$ with $\delta_i > \delta_{i-1} > 1$, the series of the calculated 
solutions:
\begin{equation*}\label{eq:Psi_vec}
\mathbf{\Psi^i} = ( y_0, \, y^i(t^i_{\delta_i}), \, y^i(t^i_{2 \delta_i}), \, \cdots, \, \underline{ y^i(t^i_{k_0 \delta_i}) }, \,  \cdots, \, y^i(T_f) )
\end{equation*}
\noindent
at each  instant $t^i_{k_i} = k_i h_i = k_{i} h_0 / \delta_i, \, k_i \in \mathcal{K}_i \subseteq  \mathcal{K}_0 \subset \mathbb{N}$ and thus $t^i_{k_0 \delta_i} = t^0_{k_0}$.
\noindent
Similarly, we denote by $\Psi^T$ the series of the points that correspond to the "true" solution:
\begin{equation*}
\mathbf{\Psi^T} = ( y_0, \, y^T(t^0_1), \, y^T(t^0_2), \, \cdots \, y^T(t^0_{k_0}), \, \, \cdots, \, y^T(T_f) )
\end{equation*}
\noindent
at each instant $t^0_{k_0}$. Note that since $h_0$ is the initial time-step and thus the "reference" time-step, $y^T$ is computed at the same instants $t_{k_0}$.
\noindent
Therefore, at each instant $t_{k_0}$, the solution is described as a finite series composed of terms that are computed  only from the power computations of the dynamic matrix (the time-step scaling factor $\delta_i$ 
allows computing each series $\mathbf{\Psi^i}$ at the same instants $t_{k_0}$ as taken for the solution $\mathbf{\Psi^0}$). At each instant $t_{k_0}$, the solution (\ref{eq:EDP_disc}) is described as 
$\mathbf{\Omega_{k0}}$-series (composed of the underlined terms of each $\mathbf{\Psi^i}, i \in \mathcal{I}$):
\begin{equation}\label{eq:omega_def}
\mathbf{\Omega_{k_0}} = ({y^0(t^0_{k_0})}, \, { y^1(t^1_{k_0 \delta_1})}, \, \cdots, \, { y^i(t^i_{k_0 \delta_i})}, \, \cdots)
\end{equation}
\noindent
with:
\begin{equation}\label{eq:omega_limit}
\lim_{i \rightarrow \infty} y^i(t^i_{k_0 \delta_i}) = \Omega^{lim}_{k_0} \approx y^T(t_{k_0})
\end{equation}
To obtain an accurate estimation of the limit $\Omega^{lim}_{k_0}$, we will describe the $\varepsilon$-algorithm, which is a convergence accelerator algorithm, whose purpose is to estimate the limit $\Omega^{lim}_{k_0}$ 
from only a few terms of $\mathbf{\Omega_{k_0}}$.
\section{Shanks transform of numerical series}
Consider $(\mathbf{S_n})_{n \in \mathbb{N}}$ (also simply $\mathbf{S_n}$), a real series that converges to a limit $S^{lim}$ for $n > n_{lim}$. To accelerate the convergence, one defines the transformation 
$\Phi$ : $\mathbb{R} \longrightarrow \mathbb{R}$, such as $\mathbf{T_n} = \Phi(\mathbf{S_n})$, whose limit is $T_{\infty}$, implies that $T_{\infty} - S^{lim} = o(S_n - S^{lim})$. In other words, we define the transformation 
$\Phi$, such as the transformed series $\mathbf{T_n}$ from the series $\mathbf{S_n}$ reaches its limit $S^{lim}$ faster than $\mathbf{S_n}$.
Among the different methods that has been established to accelerate the convergence of a series \cite{Smith} \cite{Wimp}, we consider the $\varepsilon$-algorithm \cite{Wynn} \cite{Wynn2} \cite{Graves} based on the Shanks 
transform \cite{Shanks} \cite{Singh} \cite{Sidi1} \cite{Sidi2}, that is reputed to be one the most powerful algorithm \cite{Brez2} to accelerate nonlinear series.
\subsection{The $\varepsilon$-algorithm}
Consider a series $\mathbf{S_n}$, for which we aim to estimate the limit $S^{lim}$ taking into that $S_n$ is the $n$th element of the series. From a few terms of $\mathbf{S_n}$, the limit $S^{lim}$ can be extrapolated using the 
following algorithm \cite{Wynn}:
\begin{equation}
\left\{ \begin{array}{l}
\varepsilon_{-1}^{(n)} = 0 \qquad \varepsilon_{0}^{(n)} = {S_n}, \qquad \qquad (n \in \mathbb{N}) \\
\varepsilon_{k+1}^{(n)} = \varepsilon_{k-1}^{(n)} + (\varepsilon_{k+1}^{(n)} - \varepsilon_{k-1}^{(n)})^{-1}, \qquad \qquad (k,n) \in \mathbb{N}
\end{array} \right.
\end{equation}
\noindent
where $\mathbf{S_n}$ could be a one-dimension series or a multidimensional series.
\noindent
The $\varepsilon$-algorithm can be described as a function $\mathbf{S^*_n} = S_{\varepsilon}(k,n, \mathbf{S_n})$ that extrapolates $\mathbf{S_n}$ and gives a new series $\mathbf{S^*_n}$ that converges faster to $S^{lim}$ (the limit $S^{lim}$ 
should be reached after a few terms of $\mathbf{S^*_n}$), where $\mathbf{S_n}$ is the series for which we aim to estimate the limit $S^{lim}$. The parameters of the extrapolation function $k$ and $n$ are respectively the index 
of $S_{\varepsilon}$ ($k$ is also the order of the associated Shanks transform) and the initial index of $\mathbf{S_n}$.
\noindent
Table \ref{table:Ep_algo} gives the number of terms of $\mathbf{S_n}$ and the corresponding number of computations ("nb. of cmp.") that are requested in order to compute $S_{\varepsilon}(k,n, \mathbf{S_n})$. Only even $k$ number are considered 
\cite{Wynn2} and any index $n$ from $\mathbf{S_n}$ may be considered $((k, n) \in \mathbb{N})$.
\begin{table}[!h]
\centering
\caption{\label{table:Ep_algo} Table of computations of $S_{\varepsilon}(k,n, S_n)$.}
\begin{tabular}{ccccccccccc}
\hline
$k$ & nb. of cmp. & $S_{n}$  &  $S_{n+1}$ & $S_{n+2}$ & $S_{n+3}$ & $S_{n+4}$  & $S_{n+5}$  &  $S_{n+6}$  & $S_{n+7}$  &  $S_{n+8}$  \\
\hline
\hline
0  &  1                        &   1   &  -     &  -     & -     & -   &   -  &  - & -  &  -   \\
2  &  6                        &   2  &   3	   &  1	    &  -   &  -   &  -  &  -  & -  &  -  \\
4  &  35                       &   3  &  10	   &  14	&  7   &  1	  &  -  &  -  & -  &  -  \\
6  &  204                      &   4  &  21	   & 55	    & 70   &  42  & 11	&  1	& -  &  -  \\
8  &  1189                     &  5	  &   36   & 140	&  301  & 363  &  242  &  86	 &  15	&  1	\\
\hline
\end{tabular}
\end{table}
We can deduce that the estimation of the limit $S^{lim}$ is more accurate when $k$ and $n$ are sufficiently high (depending on the convergence rate of $\mathbf{S_n}$). However, higher index $k$ involve much more computations 
that may decrease the overall efficiency of the method. As a result, a compromise has to be found between $k$ and $n$ in order to get the most accurate $S^{lim}$ with a few terms of $\mathbf{S_n}$. 
\subsection{Application to the explicit Euler scheme}
We present the general strategy applied to the explicit Euler scheme that will be transposed to the explicit-Euler based Parareal algorithm in the next section. To provide a "generic" form of the explicit Euler induction, we assume that 
$\mathbf{A}$, $\mathbf{B}$ and $u$ are constants using the "initial" time-step $h_0$. From the first recursive terms of (\ref{eq:EDP_disc}):
\begin{equation*}
\begin{array}{l}
y^0 (t^0_{1}) = (\mathbf{I} + h_0 \mathbf{A})y^0 (t_{0})  + h_0 \mathbf{B} u \\
y^0 (t^0_{2}) = (\mathbf{I} + h_0 \mathbf{A})y^0 (t^0_{1})  + h_0 \mathbf{B} u \\
\vdots \\
y^0 (t^0_{k_0})  = (\mathbf{I} + h_0 \mathbf{A}) y^{0} (t^0_{k_0-1})   + h_0 \mathbf{B} u
\end{array}
\end{equation*}
\noindent
we obtain the following definition of the explicit Euler scheme:
\begin{equation}\label{eq:Euler_exp_disc}
y^0 (t^0_{k_0})   = \left(\mathbf{I} + h_0 \mathbf{A} \right)^{k_0} y^0 (t_{0})   + \sum_{j = 0}^{{k_0}-1} \left(\mathbf{I} + \mathbf{A} \, h_0 \right)^j h_0 \mathbf{B} u,  \quad k_0 \in \mathcal{K}_0
\end{equation}
\noindent
According to the Section \ref{sec:secI}, we introduce the parameter $\delta_i$, that divides the time-step $h_0$ such as $h_i = h_0 / \delta_i$, and the time-iteration $k_i$, as a subdivision of $k_0$. Therefore, (\ref{eq:Euler_exp_disc}) is 
rewritten at each same instants:
\begin{equation}
 y^i (t^i_{k_0 \delta_i})  = \left(\mathbf{I} + \mathbf{A} \frac{h_0}{\delta_i} \right)^{\delta_i {k_i}} y^i (t_{0})  + \sum_{j = 0}^{\delta_i k_0-1} \left(\mathbf{I} + \frac{h_0}{\delta_i} \right)^j \frac{h_0}{\delta_i} 
 \mathbf{B} u,\quad k_i \in \mathcal{K}_i
\end{equation}
\noindent
For all $\delta_i$, the corresponding $\mathbf{\Omega_{k_0}}$ series ``collects'' each value of $y^i(t^i_{k_0 \delta_i})$ and one retrieve the series (\ref{eq:omega_def}):
\begin{equation*}
\mathbf{\Omega_{k_0}} = ({y^0(t^0_{k_0})}, \, { y^1(t^1_{k_0 \delta_1})}, \, \cdots, \, { y^i(t^i_{k_0 \delta_i})}, \, \cdots)
\end{equation*}
\noindent
with the limit $\Omega^{lim}_{k_0}$ given by (\ref{eq:omega_limit}). Then, for each time-iteration $k_0$, we define the {\it accelerated series} $\mathbf{\Omega^*_{k_0}}$ such as: 
$\mathbf{\Omega^*_{k_0}} = S_{\varepsilon}(k,n, \mathbf{\Omega_{k_0})}$. 
The estimated limit is noted $\Omega^{* \, lim}_{k_0}$ and is given by the last term(s) of $\mathbf{\Omega^*_{k_0}}$. 
\section{Application to the Parareal method}
\subsection{Review of the Parareal method approach}
The Parareal algorithm is a parallel method that computes numerical solutions of general systems of ordinary differential equations (ODEs) of the form (\ref{eq:EDP_fond}):
\begin{equation}\label{eq:para_sol}
 U' = f(U) \quad U(0) = U_0
\end{equation}
over the time interval $t \in [0, \, T]$ decomposed into $N$ subdivisions such as: $t = t_0, t_1, \cdots, t_j, \cdots, t_N$, where $j \in \mathcal{J} \subset \mathbb{N}$. Two propagation operators are defined:
\begin{itemize}
\item the implicit operator "coarse" $G_i(t_{j+1}, t_{j}, U_j, h_g)$, where $h_g$ is the time-step, provides a rough approximation to $U(t_{j+1})$ of the solution of (\ref{eq:para_sol}) with initial condition $U(t_{j}) = U_j$;
\item the implicit operator "fine" $F_i(t_{j+1}, t_{j}, U_j, h_f)$, where $h_f < h_g$ is the time-step, provides a more accurate approximation of $U(t_{j+1})$ of the solution of (\ref{eq:para_sol}) with initial condition $U(t_{j}) = U_j$.
\end{itemize}
The algorithm starts with an initial approximation $U_j^0, j = 0, 1, \cdots, N$ at times $t_0, t_1, \cdots t_N$ using the $G$ operator. The solution of (\ref{eq:para_sol}) given by the $G$ solver is improved by the $F$ solver. The purpose of the 
Parareal method is to integrate (\ref{eq:para_sol}) using the $G$ solver with a high time-step and then to improve iteratively, via the $F$ solver, the solution at the points calculated by the $G$ solver with a lower time-step $h_f < h_g$. 
To correct iteratively the values of the solution of (\ref{eq:para_sol}) taking into account both solutions of $G$ and $F$ solvers, (\ref{eq:Para_propag}) allows evaluating the modified points of the $G$ solver according to the iteration $k \geq 1$:
\begin{equation}\label{eq:Para_propag}
 U_{j+1}^{k+1} = G_i(t_{j+1}, t_j, U_j^{k+1}, h_g) + F_i(t_{j+1}, t_j, U_j^k, h_f) - G_i(t_{j+1}, t_j, U_j^{k}, h_g)
\end{equation}
\noindent
where $j$ designates the $j$th time subdivision of the solution of (\ref{eq:para_sol}) and $k$ the iteration during the Parareal process. The accuracy of the final solution of (\ref{eq:para_sol}) when $k \longrightarrow \infty$ depends 
on the accuracy of $F$ \cite{Vand}.
\subsection{Semi-explicit derivation of the Parareal method}
Our proposed approach follows the classical scheme of the Parareal method described in the previous subsection. 
To ensure the stability of the global algorithm, the rough solution at the first iteration $k = 1$ is performed by an implicit operator $G_i(t_{j+1}, t_{j}, U_j, h_g)$. Then, for $k > 1$,

\begin{itemize}
\item the explicit operator "coarse" $G_e(t_{j+1}, t_{j}, U_j, h_g)$, where $h_g$ is the time-step, provides a rough approximation to $U(t_{j+1})$ of the solution of (\ref{eq:para_sol}) with initial condition $U(t_{j}) = U_j$. 
\item the explicit operator "fine" $F_e(t_{j+1}, t_{j}, U_j, h_g/\delta_k)$, where $h_g/ \delta_k$ is the time-step, provides a more accurate approximation of $U(t_{j+1})$ of the solution of (\ref{eq:para_sol}) with 
initial condition $U(t_{j}) = U_j$.
\end{itemize}
\noindent
The time-step of the $F$ operator is reduced according to the $k$ iteration. Therefore, the evolution of the precision of the corrected values of the solution of (\ref{eq:para_sol}) has the same behavior than the evolution of the 
precision in the case of the simple explicit Euler method. Expressions (\ref{eq:Para_propag2a}) and (\ref{eq:Para_propag2b}) allows evaluating the modified points of the $G$ solver according to the iteration $k \geq 1$:
\begin{itemize}
\item if $k = 1$, then:
\begin{equation}\label{eq:Para_propag2a}
U_{j+1}^{k+1} = G_e(t_{j+1}, t_j, U_j^{k+1}, h_g) + F_e(t_{j+1}, t_j, U_j^k, h_f / \delta_k) - G_i(t_{j+1}, t_j, U_j^{k}, h_g)    \\
\end{equation}
\item if $k > 1$, then:
\begin{equation}\label{eq:Para_propag2b}
U_{j+1}^{k+1} = G_e(t_{j+1}, t_j, U_j^{k+1}, h_g) + F_e(t_{j+1}, t_j, U_j^k, h_f / \delta_k) - G_e(t_{j+1}, t_j, U_j^{k}, h_g)   \\
\end{equation}
\end{itemize}
\noindent
At each instant $t_{j}$, the solution is described as $\mathbf{\Omega_{j}}$-series constituted from each modified points $U_{j}^{k}$ such as:
\begin{equation}\label{eq:omega_def_para}
\mathbf{\Omega_{j}} = (U_{j}^{2}, U_{j}^{3}, \cdots U_{j}^{k}, U_{j}^{k+1}, \cdots )
\end{equation}
\noindent
with the limit $\Omega^{lim}_{j}$ given by (\ref{eq:omega_limit}). Besides the efficiency of parallelizing the operations, the main advantages of this technique is to reduce the error propagated by the $F_e$ and $G_e$ solvers thanks to the 
combination of the algorithms (\ref{eq:Para_propag2a}) and (\ref{eq:Para_propag2b}) and the solvers that manage the resolution over smaller time intervals $[t_j, \, t_{j+1}] \subset [0, \, T]$ for all $j \in \mathcal{J}$.

To evaluate the limit $\Omega^{* \, lim}_{j}$, we consider the application $\mathbf{\Omega^*_{j}} = S_{\varepsilon}(k,n, \mathbf{\Omega_{j}} + \mathbf{S^b_n})$ where we take $k = 4$ and $n = 2$ such as the calculation of the extrapolation 
is performed considering a very few terms of $\mathbf{\Omega_j}$ (according to the Table \ref{table:Ep_algo}, if $k=6$, much more computations are requested), and $\mathbf{S^b_n}$ is an alternating series that allows increasing the precision 
of the estimated limit \cite{Brez2}. This series is of the form:
\begin{equation}
S^b_n = S_b^0 + (-1)^n \sum_{j=1}^n \frac{1}{(n+1)^q}
\end{equation}
\noindent
where $S_b^0$ is the initial term of the series. Since $S^b_n$ is a convergent series that has a "resonant" transient, $q$ is a real positive number that characterizes the damping of the resonance. 
The coupling between the alternating series $\mathbf{S^b_n}$ and the series $\mathbf{\Omega_{j}}$ to accelerate, highlights an important point that concerns the resolution of the $\mathbf{\Omega_{j}}$ series. For a given $j \in \mathcal{J}$,
since only four terms of $\mathbf{\Omega_{j}}$ are requested to compute the extrapolation, an additional parameter of the extrapolation function is the distance between each term of $\mathbf{\Omega_{j}}$. We define the resolution of the series
$\mathbf{\Omega_{j}}$ as the distance $\delta^{\Omega}$-distance defined by $|\delta_{k} - \delta_{k+1}|$. For all $k$, we define $\delta_1$ and $\delta_2$ such as: $\delta_1 < |\delta_{k} - \delta_{k+1}| < \delta_2$ with 
$\delta_1 < \delta_2$ and $(\delta_1, \delta_2) \in \mathbb{R^{*+}}$.

In the next subsection, a complete procedure summarizes the computations needed to solve (\ref{eq:para_sol}) and calculate all limits $\Omega^{* \, lim}_{j}, \, j \in \mathcal{J}$.
\subsection{Optimization of the extrapolation of $\mathbf{\Omega^*_{j}}$}
To improve the use of the Wynn algorithm, an optimization procedure is considered. The goal is to adjust the damping of $\mathbf{S^b_n}$, by adjusting the coefficient $q$, in order to reduce the error 
$\mathbf{\Omega^{\varepsilon}_j}, j \in \mathcal{J}$. 
This technique implies theoretically to get a large amount of terms of the $\mathbf{\Omega_j}$-series for all $j \in \mathcal{J}$ in order to have a good estimation of the limits $\Omega^{lim}_{j}$, with which $\Omega^{* \, lim}_{j}$ will be compared. 
However, from a practically side, we consider only $\mathbf{\Omega_1}$ for the optimization and the resulting $q$ value, noted $q_{opt}$, will be propagated through the other series $\mathbf{\Omega_j}, j > 1$. 
The following steps, that initializes the optimization procedure, can be considered. 
\begin{enumerate}
\item To obtain $\Omega^{lim}_{1}$, the explicit Euler scheme is directly applied on the time-interval $[0, \, t_1]$ with a very low time-step. 
\item To obtain the optimized $q_{opt}$, the first four-term of the $\mathbf{\Omega_1}$ series are considered. The resulting $q_{opt}$ is hold for the rest of the series $\mathbf{\Omega_j}, j >1$. Therefore, the rate of 
convergence of the $\mathbf{\Omega_{j}}$ series is assumed to be constant for all $j \in \mathcal{J}$. 
\end{enumerate}
Consider the 3-variable general optimization problem (valid for all $j \in \mathcal{J}$):
\begin{equation}\label{eq:optim_omega}
\min_{\begin{array}{c} 
q > 0 \\ 
\delta_1 > 0, \, \delta_2 > 0 
\end{array}}  
\left\{ \rho S_{\varepsilon} \left( 4,2, \left( \mathbf{\Omega_{j}} + {S_b^0} + (-1)^n \sum_{j=1}^n \frac{1}{(n+1)^q} \right) \right)  - y^T(t^0_i)  \right\}  \\
\end{equation}
\begin{equation*}
 \hbox{subjected to   }  \delta_1 < |\delta_{k} - \delta_{k+1}| < \delta_2, \quad \delta_1 < \delta_2,
\end{equation*}
\noindent
where  $\rho \in \mathbb{R}^{*+}$ is a scaling factor whose purpose is to scale $\mathbf{\Omega_1}$ in order to be of the same range than the auxiliary series.
The following complete algorithmic procedure enumerates the steps in order to evaluate the limits $\Omega^{* \, lim}_{j}$ for all $j \in \mathcal{J}$.
\begin{enumerate}
 \item \underline{Parareal computation} The proposed derivation of the Parareal method solves the ODEs, using $G_i$, $G_e$ and $F_e$ solvers and applying (\ref{eq:Para_propag2a}) and (\ref{eq:Para_propag2b}) as 
 described in the previous subsection. As a result, the series $\mathbf{\Omega_j}$ are computed for all $j \in \mathcal{J}$.
 \item \underline{Initialization step} From the $\mathbf{\Omega_j}$-series, the initialization consists in calibrating the convergence accelerator, by determining firstly the $\Omega^{lim}_{1}$ using an explicit Euler scheme, and then  
 the $q_{opt}$ value for which the global error $\mathbf{\Omega^{\varepsilon}_1}$ is minimized using (\ref{eq:optim_omega}). Remind that a small index $k$ is considered to apply the Wynn algorithm that limits the number of requested terms from
 $\mathbf{\Omega_j}$ and the number of computations.
 \item \underline{Final acceleration} It follows that each approximated limit $\Omega^{* \, lim}_{j}, j > 1$ is calculated using the expression (\ref{eq:optim_omega_eval}).
\begin{equation}\label{eq:optim_omega_eval}
\Omega^{* \, lim}_{j} = S_{\varepsilon} \left( 4,2, \left( \mathbf{\Omega_{j}} + S_b^0 + (-1)^n \sum_{j=1}^n \frac{1}{(n+1)^{q_{opt}}} \right) \right)
\end{equation}
\end{enumerate}
\noindent
The series $\mathbf{\Omega^{* \, lim}}$ composed of the $\Omega^{* \, lim}_{j}$ terms for $j \in \mathcal{J}$ such as:
\begin{equation}
 \mathbf{\Omega^{* \, lim}} = ( \Omega^{* \, lim}_{1}, \, \Omega^{* \, lim}_{2}, \, \Omega^{* \, lim}_{3}, \, \cdots )
\end{equation}
constitutes the solution of (\ref{eq:para_sol}), calculated by the Parareal method at the instants $t = t_1, t_2, t_3, \cdots$ with only a few iterations, and extrapolated using the Wynn algorithm.

\begin{figure}[!h]
\centering
\includegraphics[width=13cm]{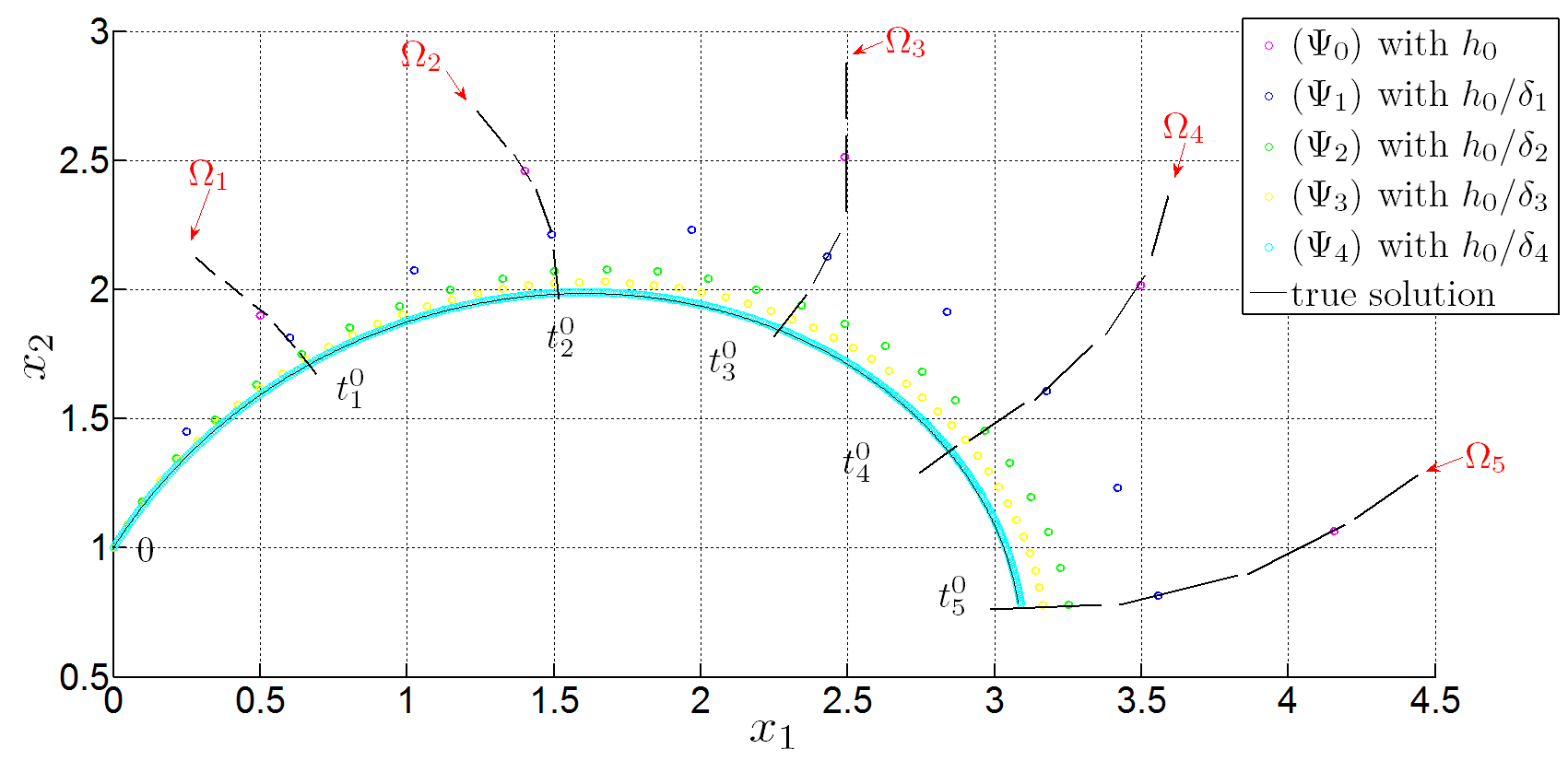}
\caption{True solution of $\Sigma$ plotted in the phase-space in comparison with explicit Euler scheme with different time-steps $h_i$ (corresponding to the $\Psi_i$ series).}
\label{fig:Euler_simple}
\end{figure}

\paragraph{Illustrative example:}
Consider in this paper, a linear second order system $\Sigma$, described by a state-space representation:
\begin{equation}\label{eq:example}
\Sigma := \left(
\begin{array}{c}
\dot{x}_1 \\
\dot{x}_2
\end{array} \right) =
 \left(\begin{array}{cc}
-1 &  5  \\
-5 &  -1  \\
\end{array} \right)
\left(  \begin{array}{c}
x_1 \\
x_2 \\
\end{array} \right) +
\left(  \begin{array}{c}
0 \\
1 \\
\end{array} \right) u
\qquad
x_1(0) = 0,
x_2(0) = 1
\end{equation}
\noindent
considering $u = 10$. Figure \ref{fig:Euler_simple} presents the different solutions of $\Sigma$, relating to different time-step in the phase-space $x_1 - x_2$ by applying the explicit Euler scheme. The "true solution" is the exact solution 
of $\Sigma$. Starting from a time-step $h_0$, that gives a very inaccurate solution, the solution is computed considering some subdivisions $\delta_i, i = 1 \cdots 5$. At each instant $t^0_{k_0}$, given some terms of the series 
$\mathbf{\Omega_{k_0}}$, the acceleration algorithm allows having an accurate estimation of $\Omega^{lim}_{k_0}$. 

\begin{figure}[!h]
\centering
\includegraphics[width=13cm]{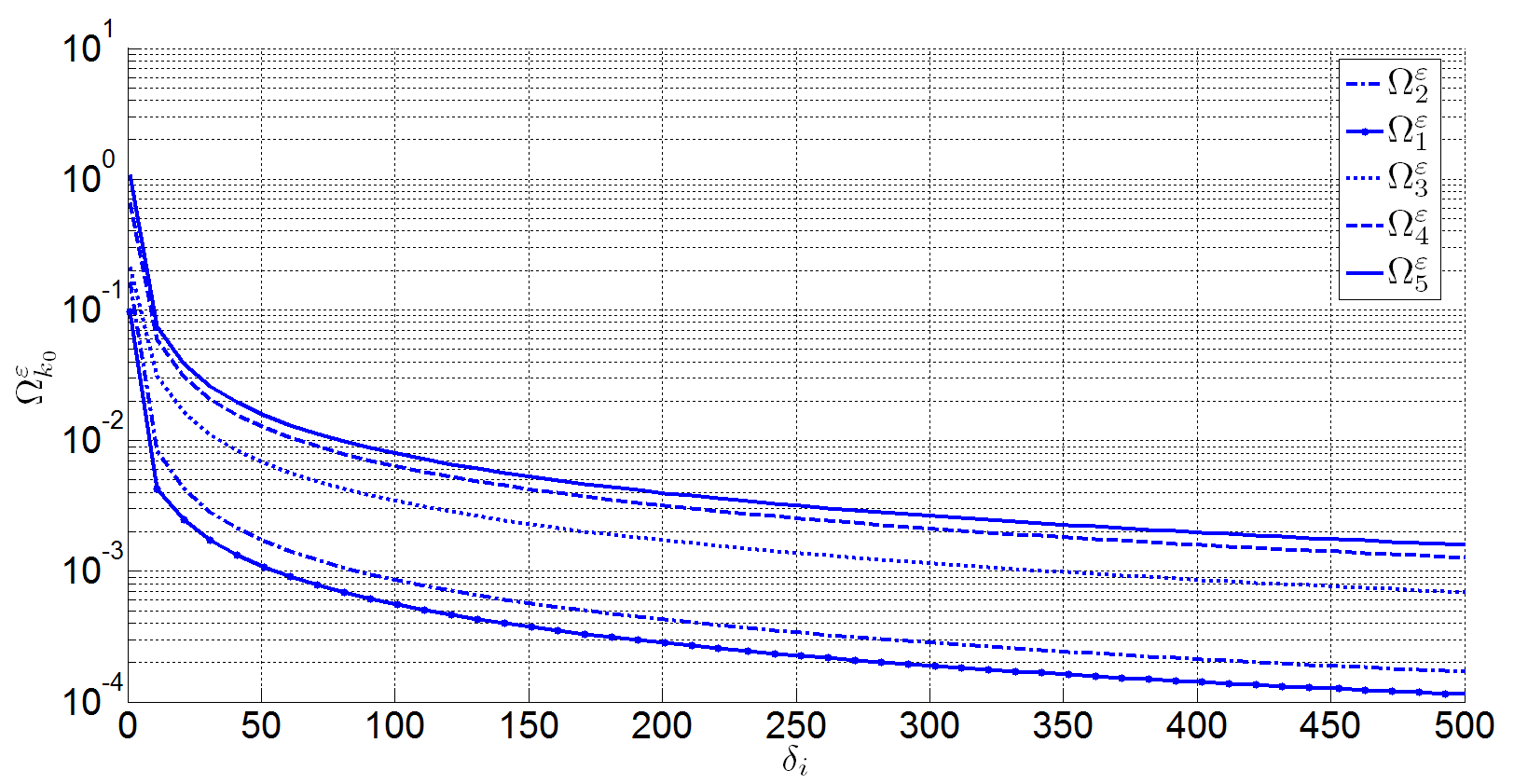}
\caption{Evolution of the $\mathbf{\Omega^*_{k_0}}$ series for $k_0 = 1 \cdots 5$ according to $\delta_i$.}
\label{fig:Omega_Euler_vec}
\end{figure}

In the case of the explicit Euler scheme, Figure \ref{fig:Omega_Euler_vec} depicts the typical logarithmic evolution \cite{Butcher} of the error $\mathbf{\Omega^{\varepsilon}_{k_0}} = |\mathbf{\Omega_{k_0}} - y^T(t^0_{k_0})|$ series 
according to $\delta_i$ for $h_0 = 0.1$ s and $k_0 = 1 \cdots 5$. 
The application of the proposed strategy to evaluate the limit of each $\mathbf{\Omega_{k_0}}$ series does not provide any significant improvement taking into account that an accurate (asymptotic) precision can be obtained usually 
from $\delta_i \geq 50$ considering small $k_0$. Indeed, the global error increases according to $k_0$ due to the error propagation throughout the integration of (\ref{eq:example}) over the time.

\begin{figure}[!h]
\centering
\includegraphics[width=13cm]{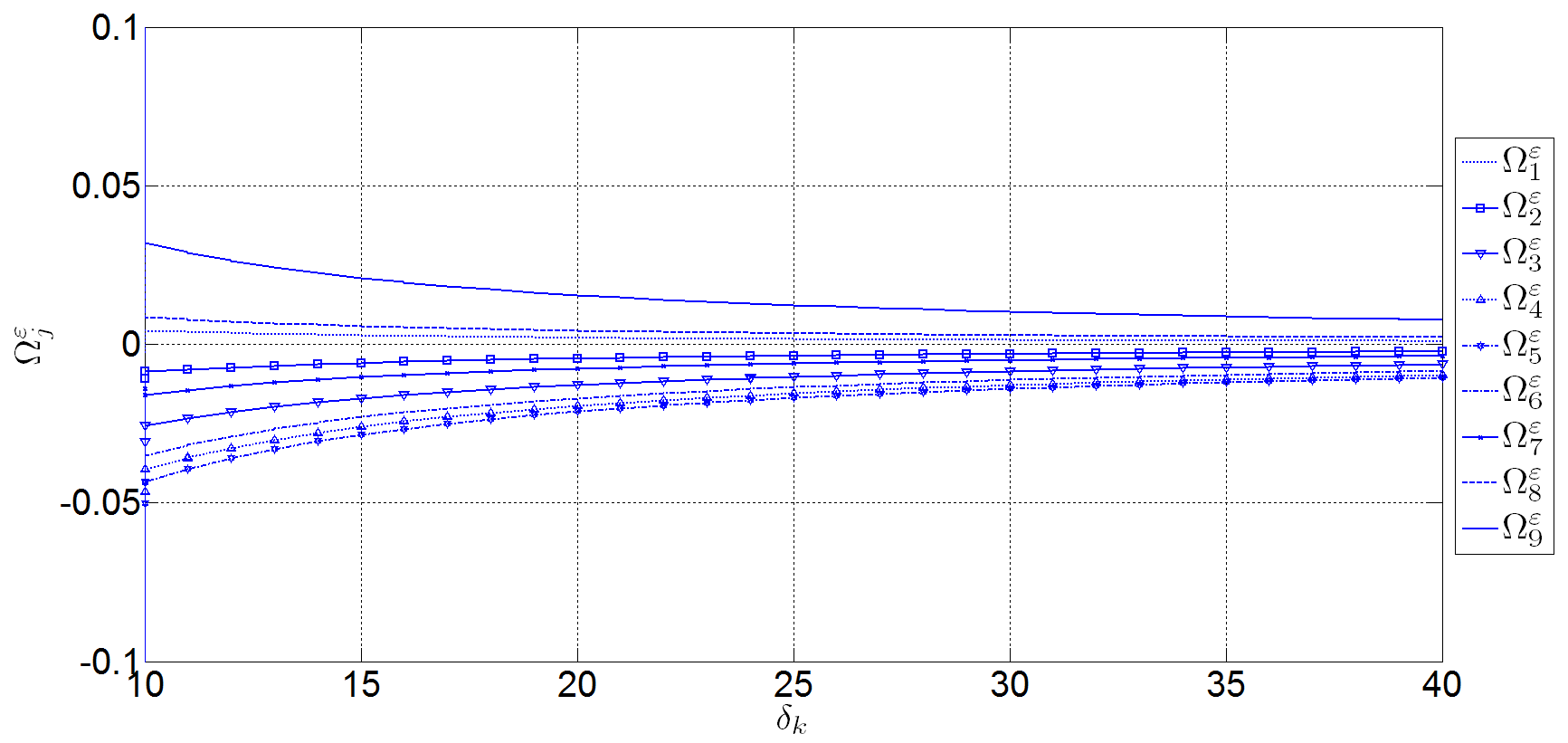}
\caption{Evolution of the $\mathbf{\Omega^{\varepsilon}_j}$ series for $j = 1 \cdots 9$ series according to $\delta_k$.}
\label{fig:Omega_Euler_vec_para}
\end{figure}

The Parareal technique is now applied to solve (\ref{eq:example}). Figure \ref{fig:Omega_Euler_vec_para} presents the evolution of each $\mathbf{\Omega^{\varepsilon}_j} = |\mathbf{\Omega_{j}} - y^T(t^0_j)|$ series according to 
$\delta_k$ for $h_0 = 0.1$ s and $j = 1 \cdots 9$.  

To verify the main assumption for which $q$ is kept constant for all $\mathbf{\Omega_j}$ (\S 4.3, item 2), we search the optimal $q_{opt}$ by applying (\ref{eq:optim_omega}) to each series $\mathbf{\Omega_j}$. 
A meta-heuristic optimization procedure, like the simulated annealing method \cite{Gran}, is applied. The following tables \ref{table:Omega_acc_1} \ref{table:Omega_acc_2} \ref{table:Omega_acc_3} \ref{table:Omega_acc_10} 
\ref{table:Omega_acc_20} \ref{table:Omega_acc_30} show the result of the search of the $q_{opt}$ values considering some different fixed $\delta^{\Omega}$-distance 
values and fixed $\delta_j$ values (the time-step divisor $h_0/\delta_i$) in the Euler scheme. For all cases, the goal is to find $q_{opt}$ such as both $|\Omega^{* \, lim}_j  -  \Omega^{lim}_j|$ and $|\Omega^{* \, lim}_j -  y(t^0_j) |$  are 
minimized separately.

\begin{table}[!h]
\centering
\caption{\label{table:Omega_acc_1} Explicit Euler scheme with $\delta_j = 500$  and  $\delta$-distance = 0.5 }
\begin{tabular}{c|cc||cc}
\hline
 $j$  &  $q_{opt} \rightarrow $ & $|\Omega^{* \, lim}_j  -  \Omega^{lim}_j| \times 10^3$   &  $q_{opt} \rightarrow $ &  $|\Omega^{* \, lim}_j -  y(t^0_j) | \times 10^3$         \\
 \hline
    1 &  0.0022  &  9.5995  &  0.0022  & 26.4484 \\
    2 &  0.0022  &  9.7795  &  0.0022 &   7.7027 \\
    3  &  0.0022 &   9.5684  &  0.0022 &   9.1885 \\
    4  &  0.0022 & 219.3907  &  0.0022 & 524.8596 \\
    5  &  0.0022 &  65.2311  &  0.0022 &   9.1266 \\
    6  &  0.0022 &   7.9504 &   0.0022 &   8.0144 \\
    7  &  0.0022 & 600.3367  &  0.0022 &   7.1330 \\
    8  &  0.0022  &  9.3852  &  0.0022 & 122.9864 \\
    9  &  0.0022 & 189.1382  &  0.0022  &  8.8399 \\
\end{tabular}
\end{table}

\begin{table}[!h]
\centering
\caption{\label{table:Omega_acc_2} Explicit Euler scheme with $\delta_j = 100$  and  $\delta$-distance = 0.1  }
\begin{tabular}{c|cc||cc}
\hline
 $j$  &  $q_{opt} \rightarrow $ & $|\Omega^{* \, lim}_j  -  \Omega^{lim}_j| \times 10^4$   &  $q_{opt} \rightarrow $ &  $|\Omega^{* \, lim}_j -  y(t^0_j) | \times 10^4$         \\
\hline
    1  &  0.0035  &  1.1811  &  $\leq 10^{-10}$  &   5.0441 \\
    2  &  $\leq 10^{-10}$  &  5.0403  &  0.0035  &  5.0051 \\
    3  &  $\leq 10^{-10}$  &  5.0403  &  0.0035  &  0.0040 \\
    4  &  0.0035  &  0.0005  &  $\leq 10^{-10}$ &    5.0441  \\
    5  &  $\leq 10^{-10}$  &  5.0403  &  0.0035  &  0.5065 \\
    6  &  $\leq 10^{-10}$  &  5.0403  &  $\leq 10^{-10}$ &   5.0441  \\
    7  &  0.0035  &  5.0359  &  0.0035 &   0.0515  \\
    8  &  $\leq 10^{-10}$  &  5.0403  &  $\leq 10^{-10}$  &   5.0441  \\
    9  &  $\leq 10^{-10}$  &  5.0403  &  $\leq 10^{-10}$  &   5.0441  \\
\end{tabular}
\end{table}

\begin{table}[!h]
\centering
\caption{\label{table:Omega_acc_3} Explicit Euler scheme with $\delta_j = 50$  and  $\delta$-distance = 0.05  }
\begin{tabular}{c|cc||cc}
\hline
 $j$  &  $q_{opt} \rightarrow $ & $|\Omega^{* \, lim}_j  -  \Omega^{lim}_j| \times 10^4$   &  $q_{opt} \rightarrow $ &  $|\Omega^{* \, lim}_j -  y(t^0_j) | \times 10^4$         \\
 \hline
    1  &  $\leq 10^{-10}$  &  5.0364 &   $\leq 10^{-10}$ &   5.0441  \\
    2  &  0.0035  &  0.0071  &  $\leq 10^{-10}$ &   5.0441  \\
    3  &  0.0035  &  0.1824  &  0.0035  &  0.6441  \\
    4  &  $\leq 10^{-10}$  &  5.0364  &  $\leq 10^{-10}$ &   5.0441  \\
    5  &  $\leq 10^{-10}$  &  5.0364  &  $\leq 10^{-10}$ &   5.0441  \\
    6  &  0.0035  &  0.6271  &  $\leq 10^{-10}$  &  5.0441  \\
    7  &  0.0035  &  0.3768  &  $\leq 10^{-10}$  &  5.0441  \\
    8  &  $\leq 10^{-10}$  &  5.0364  &  $\leq 10^{-10}$ &   5.0441  \\
    9  &  $\leq 10^{-10}$  &  5.0364  &  $\leq 10^{-10}$ &   5.0441  \\
\end{tabular}
\end{table}


\begin{table}[!h]
\centering
\caption{\label{table:Omega_acc_10} Explicit Euler scheme with $\delta_j = 500$  and  $\delta$-distance = 5   }
\begin{tabular}{c|cc||cc}
\hline
 $j$  &  $q_{opt} \rightarrow $ & $|\Omega^{* \, lim}_j  -  \Omega^{lim}_j| \times 10^4$   &  $q_{opt} \rightarrow $ &  $|\Omega^{* \, lim}_j -  y(t^0_j) | \times 10^4$         \\
\hline
    1  &  $\leq 10^{-10}$  &  5.0066  &  $\leq 10^{-10}$  &  5.0074 \\
    2  &  $\leq 10^{-10}$  &  5.0066  &  $\leq 10^{-10}$  &  5.0074 \\
    3  &  $\leq 10^{-10}$  &  5.0066  &  $\leq 10^{-10}$  &  5.0074 \\
    4  &  $\leq 10^{-10}$  &  5.0066  &  $\leq 10^{-10}$  &  5.0074 \\
    5  &  $\leq 10^{-10}$  &  5.0066  &  $\leq 10^{-10}$  &  5.0074 \\
    6  &  $\leq 10^{-10}$  &  5.0066  &  $\leq 10^{-10}$  &  5.0074 \\
    7  &  $\leq 10^{-10}$  &  5.0066  &  $\leq 10^{-10}$  &  5.0074 \\
    8  &  $\leq 10^{-10}$  &  5.0066  &  $\leq 10^{-10}$  &  5.0074 \\
    9  &  $\leq 10^{-10}$  &  5.0066  &  $\leq 10^{-10}$  &  5.0074 \\
\end{tabular}
\end{table}

\begin{table}[!h]
\centering
\caption{\label{table:Omega_acc_20} Explicit Euler scheme with $\delta_j = 100$  and  $\delta$-distance = 1 }
\begin{tabular}{c|cc||cc}
\hline
 $j$  &  $q_{opt} \rightarrow $ & $|\Omega^{* \, lim}_j  -  \Omega^{lim}_j| \times 10^4$   &  $q_{opt} \rightarrow $ &  $|\Omega^{* \, lim}_j -  y(t^0_j) | \times 10^4$         \\
\hline
    1  &  $\leq 10^{-10}$  &  4.8732  &  $\leq 10^{-10}$  &  4.8770 \\
    2  &  $\leq 10^{-10}$  &  4.8732  &  $\leq 10^{-10}$  &  4.8770 \\
    3  &  $\leq 10^{-10}$  &  4.8732  &  $\leq 10^{-10}$  &  4.8770 \\
    4  &  $\leq 10^{-10}$  &  4.8732  &  $\leq 10^{-10}$  &  4.8770 \\
    5  &  $\leq 10^{-10}$  &  4.8732  &  $\leq 10^{-10}$  &  4.8770 \\
    6  &  $\leq 10^{-10}$  &  4.8732  &  $\leq 10^{-10}$  &  4.8770 \\
    7  &  $\leq 10^{-10}$  &  4.8732  &  $\leq 10^{-10}$  &  4.8770 \\
    8  &  $\leq 10^{-10}$  &  4.8732  &  $\leq 10^{-10}$  &  4.8770 \\
    9  &  $\leq 10^{-10}$  &  4.8732  &  $\leq 10^{-10}$  &  4.8770 \\
\end{tabular}
\end{table}

\begin{table}[!h]
\centering
\caption{\label{table:Omega_acc_30} Explicit Euler scheme with $\delta_j = 50$  and  $\delta$-distance = 0.5 }
\begin{tabular}{c|cc||cc}
\hline
 $j$  &  $q_{opt} \rightarrow $ & $|\Omega^{* \, lim}_j  -  \Omega^{lim}_j| \times 10^4$   &  $q_{opt} \rightarrow $ &  $|\Omega^{* \, lim}_j -  y(t^0_j) | \times 10^4$         \\
\hline
    1 &  0.0022  &  8.7869  &  0.0022  &  6.9852 \\
    2  &  0.0022  & 60.2327  &  0.0022  &  8.2166 \\
    3  &  0.0022  &  8.9733  &  0.0022 & 863.4299 \\
    4  &  0.0022  & 18.8258  &  0.0022  & 69.9347 \\
    5  &  0.0022  &  8.0245  &  0.0022  &  9.9169 \\
    6  &  0.0022  &  9.8895  &  0.0022  &  9.7054 \\
    7  &  0.0022 & 787.8841  &  0.0022  &  9.8428 \\
    8  &  0.0022 &   8.3683  &  0.0022 & 353.2981 \\
    9  &  0.0022 &   9.3224  &  0.0022  &  9.8790 \\
\end{tabular}
\end{table}
The estimated limit $\Omega^{* \, lim}_j$ is compared to the limit $\Omega^{lim}_j$ given by the asymptotic behavior of $\mathbf{\Omega_j}$ (via the Euler scheme) and the true solution $y(t^0_j)$. 
The results show that a better estimation of the limit $\Omega^{* \, lim}_j$ is obviously obtained for a high value of $\delta_i$ but in other hand, the corresponding $\delta^{\Omega}$-distance has to be properly bounded. In particular, we deduce a 
rough approximation of the lower bound such as $\delta_1 = 0.5$. Table \ref{table:Omega_acc_20} presents the results for which the best estimations of the limits have been obtained. These results confirm that the same $q_{opt}$ can be used to 
treat all $\Omega_j, j \in \mathcal{J}$.

\paragraph{Remark} In the case where the system (\ref{eq:para_sol}) has to be solved over a large time-interval, the initialization step can be applied periodically to "refresh" the optimized parameters and thus to ensure good performances.
\section{Conclusion} We presented a derivation of the Parareal algorithm scheme that uses a convergence accelerator algorithm in order to improve the precision of the resulting solution at each instant. The main improvements concern the use of an 
explicit solver for the $G$ and $F$ operators and the use of an acceleration technique, like the Wynn algorithm, to improve the convergence. The proposed method has the following features:
\begin{itemize}
\item The computations are based essentially on the power computation of the dynamic matrix $(\mathbf{I} + \mathbf{A_{k_0}} h_i)$.
\item Since the power computations involve several time-steps (the initial time-step $h_0$, and the different division $\delta_k$ requested by the $\varepsilon$-algorithm), these computations may be parallelized.
\item Multidimensional ODEs can be considered.
\end{itemize}
Future investigations include the improvement of the series accelerator algorithm relating to the alternating series, the development of an adaptive scheme (e.g. a variable time-step scheme), the development of a symplectic approach and the 
extension of the method to solve DAEs / stiff systems. Moreover, the association and implementation with existing innovative methods are also of interest. A demonstration code is available upon request to the author.

\end{document}